# The automation of optical transient discovery and classification in Rubin-era time-domain astronomy

Nabeel Rehemtulla ⓘ [1,2,3] ✉, Michael W. Coughlin ⓘ [4,5], Adam A. Miller ⓘ [1,2,3] & Theophile Jegou du Laz[5,6]

Robotic wide-field time-domain surveys, such as the Zwicky Transient Facility and the Asteroid Terrestrial-impact Last Alert System, capture dozens of transients each night. The workflows for discovering and classifying transients in survey data streams have become increasingly automated over decades of development. The recent integration of machine learning and artificial intelligence tools has produced major milestones, including the fully automated end-to-end discovery and classification of an optical transient, and has enabled automated rapid-response space-based follow-up. The now-operational Vera C. Rubin Observatory and its Legacy Survey of Space and Time are accelerating the rate of transient discovery and producing large volumes of data at incredible rates. Given the expected order-of-magnitude increase in transient discoveries, one promising path forwards for optical time-domain astronomy is heavily investing in accelerating the automation of our workflows. Here we review the current paradigm of real-time transient workflows, project their evolution during the Rubin era and present recommendations for accelerating transient astronomy with automation.

Most modern wide-field time-domain surveys, including the Zwicky Transient Facility (ZTF)[1], the Asteroid Terrestrial-impact Last Alert System (ATLAS)[2,3] and the Legacy Survey of Space and Time (LSST)[4], adopt image differencing to identify new, moving or varying sources. This involves subtracting a reference co-addition from each new image, typically over a fixed grid of sky fields. Residual pixel values in the resulting difference image, usually to the 5σ threshold, trigger 'alerts' that indicate changes relative to the reference image. 'Bogus' alerts arise from artefacts of the image subtraction process, rather than astrophysical sources[5], and have thus far been an inevitable nuisance of image differencing. They are handled well by machine learning (ML) classifiers, which have demonstrated greater than 99% accuracy in separating real and bogus alerts[5–11]. Although one could call real/bogus a 'solved problem', there is still demand for advancements. Researchers typically invest considerable time creating a bespoke large, labelled real/bogus training set and model for each new survey, but adopting domain adaptation techniques can alleviate this[12–15]. The efficiency offered by these techniques is especially valuable in the early stages of a survey; that is, when little (or no) labelled data exist, and when there are many image-differencing surveys needing real/bogus models. Both cases are pertinent, given the recent or imminent commencement of LSST, the Nancy Grace Roman Space Telescope[16] and numerous supporting surveys, such as the La Silla Schmidt Southern Survey[17].

Alerts are relayed as an 'alert stream' to 'alert brokers', which process, augment, filter and serve alerts to researchers. Alert packets contain information computed from the image (such as the object brightness and sky position), as well as supplemental information (for example, ML scores and external catalogue cross-matches). No two brokers are identical, so one can express a preference by choosing which broker(s) to use. The Automatic Learning for the Rapid Classification of Events (ALeRCE)[18] and Fink[19] brokers offer ML models tailored to specific science cases[20], whereas the Arizona–NOIRLab Temporal Analysis and Response to Events System (ANTARES) broker[21] empowers users with the ability to define fully custom alert filters. LSST produces an order of magnitude more alerts per night than previous-generation surveys like ZTF: ~$10^7$ versus $10^5$–$10^6$ alerts per night. This weighty alert stream is relayed to seven pre-approved brokers[18,19,21–23]. Specializing brokers to specific utilities or science cases would be effective for reducing overlapping work; at present, many brokers offer very similar, sometimes redundant, functionalities. The multitude of surveys producing rich alert streams is a novel challenge in the Rubin era, and brokers will need to develop new functionality to support seamless syntheses of these data for easy downstream use by researchers.

[1]Department of Physics and Astronomy, Northwestern University, Evanston, IL, USA. [2]Center for Interdisciplinary Exploration and Research in Astrophysics (CIERA), Evanston, IL, USA. [3]NSF-Simons AI Institute for the Sky (SkAI), Chicago, IL, USA. [4]School of Physics and Astronomy, University of Minnesota, Minneapolis, Minnesota, USA. [5]NSF Institute on Accelerated AI Algorithms for Data-Driven Discovery (A3D3), https://a3d3.ai/. [6]Division of Physics, California Institute of Technology, Pasadena, CA, USA. ✉e-mail: nabeelr@u.northwestern.edu

After being enriched with information by an alert broker, considerable filtering of alerts is necessary to select only those alerts that are related to one's astrophysical phenomena of interest. Alert filters are a set of static conditions on the contents of the alert packets and are mostly custom to a particular science case. Alert filter components can include, for example, rejecting alerts from observations with poor seeing, requiring that the alert is not coincident with a catalogued star[24] or requiring that the alert implies a rapid light curve decline rate[25]. Sources producing alerts that pass a given filter are known as 'candidates'. These candidates are presented to researchers through an interface with the alert broker, often a convenient web-based interface for viewing data like those provided by ALeRCE[18], Lasair[23,26], ANTARES[21] and Fink[19]. Some teams use more feature-rich platforms called target and observation managers (TOMs)[27] or 'marshals' like YSE-PZ[28] and SkyPortal[29,30]. These TOMs/marshals integrate additional capabilities such as scheduling and optimizing new observations, conducting data analysis, hosting discussions and more. It is crucial for researchers to carefully construct alert filters and their components to minimize selection biases and maximize the broker's computational efficiency. These considerations are especially important as LSST is substantially expanding data volumes and growing discovery spaces beyond the reach of existing catalogues.

Researchers visually inspect (or 'scan') candidates to separate genuine sources of interest from false-positive candidates. Scanning involves considering information in image cutouts, light curves, alert packet metadata, external catalogues and more. Sources found while scanning are often reported to the transient name server (TNS), the official International Astronomical Union mechanism for reporting transients to the community. While scanning has historically been effective, it becomes increasingly, and eventually prohibitively, costly for large volumes of candidates or very time-sensitive candidates. Many specialized ML tools have been developed to ease and accelerate scanning by automating it. The Finding Luminous and Exotic Extragalactic Transients models (FLEET)[31,32], tdescore[33], the NEural Engine for Discovering Luminous Events (NEEDLE)[34], the ATLAS Virtual Research Assistant[35], and BTSbot[36] have all assisted researchers with identifying transients of interest in real survey data. The Gravitational-wave Optical Transient Observer[37] has also demonstrated that crowd-sourcing scanning via citizen scientists is viable for easing scanning loads[38].

As is the case for real/bogus models, the ML in these automated scanning utilities can often be simplistic and brittle. This introduces weaknesses such as preventing transfer to new surveys, failing when input data are incomplete, requiring large quantities of labelled data and requiring costly time investments for model design. Researchers creating such utilities should familiarize themselves with the best practices in the relevant ML literature, rather than the astronomy literature alone, to leverage the latest advances from the continually rapid progress in ML and artificial intelligence (AI).

For sources of sufficient interest, follow-up observations are requested to characterize the transient. Spectroscopic resources are not expected to grow at the same rate as transient discovery, so the fraction of transients with spectroscopic classifications will plummet in the Rubin era. It will therefore be more critical than ever to maximize the scientific return from one's spectroscopic time.

Automating the triggering of follow-up boosts one's efficiency by saving time and minimizing workflow latency. This practice has long been commonplace in gamma-ray burst astronomy[39] and, despite critical challenges, is becoming increasingly prevalent in supernova astronomy. In the case of supernovae, automated follow-up must be paired with automated transient identification to best realize their benefits. Given that follow-up resources are scarce, purity (the fraction of sources selected for follow-up that are genuinely of interest) must be a priority. Numerous factors have contributed to the recently increasing prevalence of automated transient identification and follow-up including: (1) the roboticization of observing facilities; (2) support for automated triggering by marshals/TOMs; and (3) the development of (usually ML-based) high-purity automated scanning tools. All of these factors, however, are limited to particular cases and generalized solutions are far from being realized.

The Palomar 60-inch telescope (P60) is a notable case study for the successful deployment of automated follow-up utilities into production workflows, as it has been at the heart of numerous such programmes over the years. P60 has been configured to automatically respond to gamma-ray bursts[40], bright supernovae[36,41] and nearby infant supernovae[42] following automated triggers from ML models and discoveries by the Neil Gehrels Swift Observatory (Swift)[43]. These programmes exemplify the outsized impact that small-aperture telescopes equipped with low-resolution spectrographs often have on automation-forward programmes. Tools prototyped on these small-aperture facilities are now ready to contribute to the transient and host galaxy programmes[44,45] at larger-aperture facilities equipped with massively multiplexed fibre-fed spectrographs such as the Dark Energy Spectroscopic Instrument (DESI)[46] and the 4-metre Multi-Object Spectrograph Telescope (4MOST)[47]. Through these new spectrographs, automated transient workflows will support the immense spectroscopic samples that will drive state-of-the-art studies of transient demographics, Type Ia supernova cosmology and more.

Even with facilities like DESI and 4MOST, the vast majority of transients detected by LSST will never receive spectroscopic observations. Substantial effort has been invested in photometric transient classification to maximize the utility of these transients[48–53]. Transient samples produced by such tools are already pure enough to contribute to even very sensitive cosmological analyses[54,55]. The field of photometric transient classification arrived at this point thanks in large part to a stimulus from the Photometric LSST Astronomical Time-Series Classification Challenge (PLAsTiCC)[56,57] and its extension ELAsTiCC, which created large datasets of mock LSST light curves[58]. The vast majority of photometric classification studies operate exclusively on simulated data like PLAsTiCC and ELAsTiCC and do not consider various idiosyncrasies of real-time data streams. For the most part, the same is true of anomaly detection studies, which aim to fulfil one of the most exciting prospects for Rubin-era transient astronomy: the discovery of entirely new classes of transient events. Researchers aiming to produce photometric classifiers and anomaly detectors viable for real-time deployment must outgrow using only mock data and face the messiness of real data and real-time data streams. The areas of photometric classification and anomaly detection have suffered from the lack of standardized benchmarks. For the most part, each new study presents slightly different performance metrics on, at best, slightly different datasets, so fair comparisons of models are rarely possible. Establishing such benchmarks will elucidate the relative performance of the many available model architectures and thus support the creation of maximally effective tools[59].

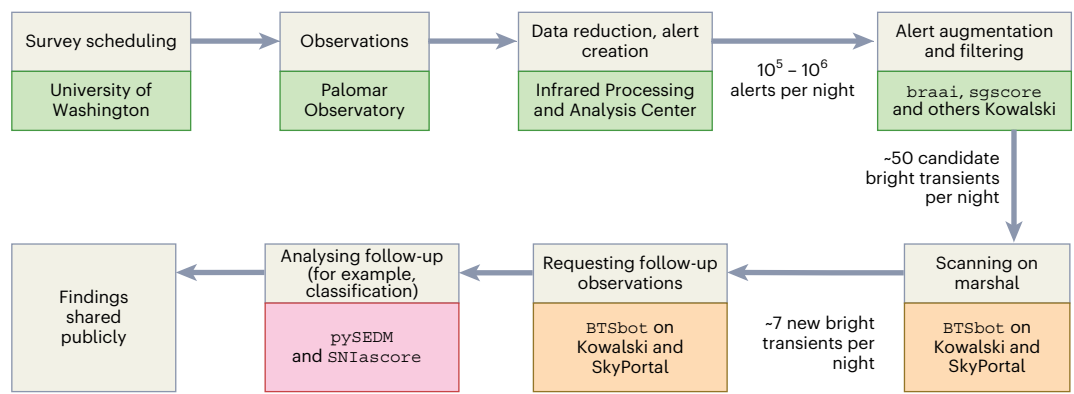

**Fig. 1 | The fully automated BTS discovery and classification workflow.** Green boxes mark highly automated stages, rarely requiring human intervention. Orange boxes denote stages only automated for the bright transients in consideration by BTS. The pink box represents a stage that often requires human intervention. The typical nightly numbers of alerts or sources remaining in each stage are also shown.

## Case study on the fully automated ZTF Bright Transient Survey workflow

We introduce the ZTF Bright Transient Survey (BTS)[36,60,61] as a case study on the trials faced when creating highly automated workflows. BTS has realized a fully automated end-to-end supernova discovery to classification workflow. Following Fig. 1, this begins with infrastructure general to ZTF, such as scheduling[62], observations, data reduction and alert generation[63]. Alerts are then relayed to brokers including Kowalski[9], which augments alerts with ML scores (for example, from braai)[9,64] and catalogue cross-matches (for example, the star/galaxy catalog from sgscore)[24] and feeds the ZTF public alert stream[65] through the BTS alert filter[61], which thins the alert stream down to transient candidates. Sources passing the alert filter are sent to SkyPortal[29,30] which checks if BTSbot[36] criteria are satisfied. When they are, candidates are catalogued as bona fide bright transients, reported to TNS and sent to the SED Machine (SEDM)[66–68] spectrograph on P60-all using SkyPortal. SEDM then conducts its observations robotically. The resulting data are automatically reduced at the P60 by pySEDM[69] and fed into SNIascore[70], running on the P60 computer at Caltech campus. If SNIascore returns a confident prediction, the P60 computer sends the source's classification as a type Ia supernova to SkyPortal and posts it to TNS.

Supernova 2023tyk did not require any human action from its discovery to its spectroscopic classification being posted to TNS, and BTS now regularly achieves fully automated discoveries and classifications of this sort: >200 type Ia supernovae since supernova 2023tyk have been triggered on by BTSbot and classified by SNIascore. This seamless end-to-end automation was enabled by a vision for automation shared across teams and hierarchies, by astronomy researchers specializing in ML/AI and by research software engineers dedicated to building complex infrastructure. No single group could have delivered this product, rather, many groups worked together to enable this. More than five years elapsed between BTS's commencement and this milestone being met, but these lessons chart a course for more generalized, quicker adoption of automation in workflows.

## Rapid-response follow-up as a rising application for automation

Extremely low-latency follow-up and classifications are required to open the world of studying cosmic explosions as they begin. This is of particular importance for multi-messenger astronomy, where robotic follow-up is required to take advantage of the sub-minute latency now enabled by real-time alert systems produced by gravitational-wave detector networks[71], for which rates of detections are rapidly increasing[72]. For events like binary neutron star mergers, researchers often receive phone calls to get them to their computers as quickly as possible to trigger follow-up observations. This work, which can happen at any time of any day, is extraordinarily disruptive. Therefore, automated, optimized and robotic observations of transients, as enabled by the workflows described here, are essential for maintaining the enthusiasm for follow-up—especially in areas like gravitational-wave counterparts, which involve rare transients with high false alert rates. While many of the scheduling tools that enable follow-up are now battle-hardened after years of development[73], the existence of fully automated candidate vetting workflows and their communication to the broader community remains both a technical and sociological challenge that must be faced to take advantage of the improving multi-messenger instruments[74], although tools like the Treasure Map[75] are a good start.

Automated, rapid-response follow-up pipelines have also begun to advance studies of nearby, infant supernovae. The Distance Less Than 40 Mpc Survey (DLT40)[76] and ZTF have both created and demonstrated such pipelines on numerous events, creating remarkably comprehensive and informative datasets[42,77]. The ZTF pipeline, called BTSbot-nearby, and Swift have also pioneered rapid-response space-based follow-up: Swift observations of supernova 2025qtt began just 6 minutes after its automated discovery and request for follow-up by BTSbot-nearby[78]. In this instance the automated workflows provide a clearer view of the earliest phases of supernovae, helping us understand the supernova physics and the link between the supernova and its progenitor system. These systems also affirm the importance of equipping observing facilities, new and old, with software to support such follow-up.

There are numerous facilities that support programmatically submitted observations including Swift[79], such as those in the Astrophysical Events Observatories Network (AEON; notably including the SOAR and Gemini telescopes)[80], those in the Las Cumbres Observatory network[81], the Liverpool Telescope[82], the MMT telescope and the GROWTH-India Telescope[83]. Compatibility in these pipelines with established standards, such as those maintained by the International Virtual Observatory Alliance, are important for managing complexity and interoperability. To this end, AEON facilities have prioritized establishing a common observations request language. Still, few facilities allow programmatically submitted time-of-opportunity requests, which are critical for achieving the lowest latencies and probing transients at phases that are nearly impossible to access with traditional methods.

## Discussion and conclusions

We have described the current paradigm of transient discovery and classification and presented our views for how this paradigm should evolve to maximize the return of LSST and other new/upcoming surveys. Automation already plays an indispensable role in transient workflows, and it has long been expected that more widespread automation would be compulsory for discovery in the future of optical time-domain astronomy (TDA)[5,6]. In particular, this involves supplanting the role of humans in the real-time decision loops with autonomous data-driven methods. There is no single action or set of actions that any one group can take to broadly enable this. Instead, all parties involved from survey management to astronomy researchers and funding agencies need to invest in this vision. Efforts thus far, such as those from ZTF and DLT40, have been relatively narrow in scope and time consuming to implement. Further development, generalization and support for automation will enhance and extend the knowledge that can be extracted from LSST.

The deployment of ML/AI and automation tools into real-time production environments is nuanced and challenging, while also central to realizing this vision. It requires deep expertise in the technologies adopted within a given alert broker, and astronomers are rarely trained in these skill sets. Thus, the engagement of research software engineers and computational scientists is critical for advancing observational transient astronomy. The ALeRCE broker is a prime example of this: multiple full-time engineers and a standardized model development pipeline—including quality assurance and consistency checks—have enabled the deployment of numerous ML models applied to real-time ZTF data. Despite their undeniable value, there are very few concrete pathways for astrophysics-related data and software engineering jobs with long-term stability. These jobs sometimes exist and are sometimes continuously funded in certain labs and well-funded institutions but remain incredibly rare overall. The LSST Discovery Alliance's LINCC programme employs software engineers, recognizing the importance of software development to Rubin-era TDA, but similar roles still remain scarce.

Fair, representative and extensive evaluation is a necessity for building trust in novel ML/AI tools intended for application to real-time workflows. Established methods in TDA have generally been successful in advancing our understanding of time-domain phenomena, so ML/AI practitioners must conceptualize, evaluate and present their models with the utmost rigour[84] to demonstrate their advancement over baselines and gain adoption of their tools. Reproducibility is key to robust science but is often forgotten in astro-ML studies; making code bases and training sets easily available is simple step to support this. Benchmarks that standardize datasets and evaluation metrics are rare but can play a critical role in making the growing world of ML models for transient astronomy easier to parse. Finally, evaluation on data unseen during training is good, but there is no substitute for operating on real-time data streams, where the strengths and weaknesses of a tool can be most completely uncovered.

With the Rubin era now upon us, it is clear that rapid-response modes for large aperture telescopes will be essential to maximize transient astronomy. A wider adoption of queue-based observing and support for programmatically submitted requests—especially time-of-opportunity requests—would facilitate more transient science without great disruption to other users. To give assurance to these projects that such low-latency responses will not yield wasted telescope time, it is also important to maximize the purity of these triggers, both in the case of optical TDA and the multi-messenger instruments that sometimes cause them. Finally, continued support for data science training and software infrastructure, particularly for projects with commitments to open-source development, are key to producing the best possible outcomes for astronomers and astronomy.

## Data availability
The data used to write this Perspective are available from the corresponding author upon request.

## Acknowledgements
We thank E. C. Bellm and Y. (Vic) Dong for their thoughtful comments. The ideas presented here have been improved through discussions at the Astroinformatics 2024 conference in Patagonia, Chile, and the 2025 Foundation Models for Astronomy Workshop at the Center for Computational Astrophysics of the Flatiron Institute. We thank the respective organizers and sponsors for providing rich discussion environments. Zwicky Transient Facility access for N.R. and A.A.M. was supported by Northwestern University and the Center for Interdisciplinary Exploration and Research in Astrophysics (CIERA). A.A.M. is supported by DoE award no. DE-SC0025599. A.A.M. is also supported by Cottrell Scholar award no. CS-CSA-2025-059 from the Research Corporation for Science Advancement. N.R. is supported by NSF award no. 2421845 and a Northwestern University Presidential Fellowship award. M.W.C. acknowledges support from the National Science Foundation under grant nos. PHY-2308862 and PHY-2117997. Based on observations obtained with the Samuel Oschin Telescope 48-inch and the 60-inch Telescope at the Palomar Observatory as part of the Zwicky Transient Facility project. ZTF is supported by the National Science Foundation under grant nos. AST-1440341 and AST-2034437 and currently award no. 2407588. ZTF receives additional funding from the ZTF partnership. Current members include Caltech, USA; Caltech/IPAC, USA; University of Maryland, USA; University of California, Berkeley, USA; University of Wisconsin at Milwaukee, USA; Cornell University, USA; Drexel University, USA; University of North Carolina at Chapel Hill, USA; Institute of Science and Technology, Austria; National Central University, Taiwan, and OKC, University of Stockholm, Sweden. Operations are conducted by Caltech's Optical Observatory (COO), Caltech/IPAC and the University of Washington at Seattle, USA. SED Machine is based upon work supported by the National Science Foundation under grant no. 1106171. The Gordon and Betty Moore Foundation, through both the Data-Driven Investigator Program and a dedicated grant, provided critical funding for SkyPortal. This research has made use of NASA's Astrophysics Data System.

## Author contributions
N.R. led the writing effort for all sections. M.W.C. wrote text in the main body and led the discussion of automation for multi-messenger astronomy. A.A.M. supported writing and copy-editing in all sections. T.J.d.L. provided consultation for technical details, wrote text in the main body and copy-edited. All authors jointly contributed to the conceptualizing the ideas presented here.

## Competing interests
The authors declare no competing interests.

## Additional information
**Correspondence** should be addressed to Nabeel Rehemtulla.

**Peer review information** *Nature Astronomy* thanks Julien Peloton and the other, anonymous, reviewer(s) for their contribution to the peer review of this work.

**Reprints and permissions information** is available at www.nature.com/reprints.

**Publisher's note** Springer Nature remains neutral with regard to jurisdictional claims in published maps and institutional affiliations.